# Cybersecurity in Transportation Systems: Policies and Technology Directions

**Ostonya Thomas**
Ph.D. Student, Glenn Department of Civil Engineering
Clemson University, Clemson, South Carolina, 29634
Email: ostonyt@clemson.edu

**M Sabbir Salek, Ph.D.**
Adjunct Faculty, Glenn Department of Civil Engineering
Clemson University, Clemson, South Carolina, 29634
Email: msalek@clemson.edu

**Jean-Michel Tine**
Ph.D. Student, Glenn Department of Civil Engineering
Clemson University, Clemson, South Carolina, 29634
Email: jtine@clemson.edu

**Mizanur Rahman, Ph.D.**
Assistant Professor, Department of Civil, Construction & Environmental Engineering
The University of Alabama, Tuscaloosa, Alabama, 35487
Email: mizan.rahman@ua.edu

**Trayce Hockstad, J.D., M.A.**
Law & Policy Analyst, Transportation Policy Research Center
The University of Alabama, Tuscaloosa, Alabama, 35487
Email: tahockstad@ua.edu

**Mashrur Chowdhury, Ph.D., P.E.**
Eugene Douglas Mays Chair of Transportation, Glenn Department of Civil Engineering
Clemson University, Clemson, South Carolina, 29634
Email: mac@clemson.edu

**ABSTRACT**
The transportation industry is experiencing vast digitalization as a plethora of technologies are being implemented to improve efficiency, functionality, and safety. Although technological advancements bring many benefits to transportation, integrating cyberspace across transportation sectors has introduced new and deliberate cyber threats. In the past, public agencies assumed digital infrastructure was secured since its vulnerabilities were unknown to adversaries. However, with the expansion of cyberspace, this assumption has become invalid. With the rapid advancement of wireless technologies, transportation systems are increasingly interconnected with both transportation and non-transportation networks in an internet-of-things ecosystem, expanding cyberspace in transportation and increasing threats and vulnerabilities. This study investigates some prominent reasons for the increase in cyber vulnerabilities in transportation. In addition, this study presents various collaborative strategies among stakeholders that could help improve cybersecurity in the transportation industry. These strategies address programmatic and policy aspects and suggest avenues for technological research and development. The latter highlights opportunities for future research to enhance the cybersecurity of transportation systems and infrastructure by leveraging hybrid approaches and emerging technologies.

**Keywords:** Cybersecurity, Cyber incidents, Transportation infrastructure security, and Emerging technologies for cybersecurity.

## INTRODUCTION

As society advances and demands on infrastructure increase, various sectors aim to boost efficiency through digitalization. Cybersecurity has been a concern since the rise of the internet, digitalization, and automation. In recent years, privacy violations and major cyberattacks have surged unprecedentedly. An IBM annual report revealed that the average cost of a data breach in the U.S. in 2022 was about US $9.44 million (*1*). In 2021, numerous hospitals, schools, and municipal governments in the U.S. experienced shutdowns due to cyber attacks (*2*). Additionally, the Colonial Pipeline breach led to a widespread panic, significantly impacting consumers (*3*). A survey indicated that weekly ransomware attacks in the transportation sector surged by 186% between June 2020 and June 2021 (*4*).

The transportation sector is rapidly transforming digitally to improve efficiency, functionality, and safety. While this digitalization offers benefits, bad actors exploit the expanded cyberspace in transportation, causing widespread damage (*5*). Cyber attacks have affected multiple transportation modes globally (*6*), highlighting the critical need for robust cybersecurity measures in multimodal transportation systems. Customers and employees in the industry have had data compromised due to cyber attacks (*6-10*). With the global average cost of a data breach in transportation surpassing US $4 million in 2023 (*11*), the importance of cybersecurity cannot be overstated. The industry, including aviation, road, rail, and maritime sectors, has become increasingly vulnerable to cyber attacks, jeopardizing critical infrastructure security. For example, a 900% rise in maritime cyber attacks from 2017 to 2021 (*12*) underscores the proliferation of vulnerabilities. This escalation reflects the evolving interconnected systems and reliance on information technologies, leading to a more complex cyber threat environment (*13*). Understanding the reasons for the surge in cybersecurity vulnerabilities is crucial to developing appropriate counterstrategies.

Cyber attacks on transportation systems, including in aviation, maritime, rail, and road transportation, have continued to occur. In 2022, Accelya, a technology firm serving over 250 airlines, was affected by ransomware, compromising company data (*6, 14*). This incident highlighted the critical importance of supply chain cybersecurity in the transportation industry. On June 18th, 2021, falsified automatic identification system (AIS) data indicated that two North Atlantic Treaty Organization (NATO) warships sailed from Odessa to Sevastopol, despite evidence showing the ships never left Odessa (*6, 7, 15*). This incident demonstrated the vulnerability of AISs to cyber attacks like spoofing, emphasizing the risks of digital infrastructure with known weaknesses. The first known double extortion ransomware attack in the US freight and rail industry occurred in 2021 when OmniTRAX was targeted, leading to the leak of 70 GB of data (*6, 16*). This incident underscored the rail sector's vulnerability to cybercriminals capable of locking them out of their own data and infrastructure. In 2022, hacktivists disabled Novosibirsk's traffic management system in Russia, preventing traffic engineers from detecting heavy traffic flow, and impacting bus, trolley, and taxi services (*6, 17*). It illustrated how an attack on digital transportation infrastructure can lead to significant disruptions and inconvenience for commuters. In June 2023, the Oregon Department of Motor Vehicles was affected by a global hack targeting the MOVEit file transfer system, compromising the information of active Oregon driver's licenses, permits, and IDs (*18*). These incidents highlight the growing threats to digital transportation infrastructure and the urgent need for comprehensive cybersecurity measures to safeguard public safety, operations and consumer interests.

This paper focuses on overarching trends and strategies applicable to all transportation modes. It identifies cybersecurity challenges in multimodal systems, examines the rise in vulnerabilities, and proposes programmatic, policy, and technological solutions to enhance industry-wide cybersecurity.

## WHY CYBER VULNERABILITIES ARE INCREASING

The cyber threat landscape in aviation, road, rail, and maritime transportation has grown increasingly complex due to interconnectivity and reliance on information and automation technologies (*13*). Advancements in technology have expanded the attack surface in transportation. This section examines the causes behind this rise in cyber vulnerabilities.

**Increased Connectivity and Automation in Transportation Systems**
Transportation systems have grown in size and complexity. As a result, intelligent transportation systems (ITS) and intelligent management systems (IMS) technologies are now essential. Connectivity enables efficient system communication and automation minimizes human interaction. These technologies improve efficiency, functionality, and safety but also increase the risk of cyber attacks. Two key drivers of cyber attacks are the increased reliance on internet-of-things (IoT) technologies in transportation and the emergence of connected and autonomous vehicles (CAV).

*Increased Reliance on IoT Technologies*
The IoT can be understood as a network of devices which can communicate with each other. IoT technologies have become increasingly important in transportation, from CAV applications to managing information and operational systems. Demertzi et al. highlighted that IoT devices can be vulnerable to cyberattacks due to limited computing power and hardware limitations, which hinder security mechanisms (*19*).

*Connected Autonomous Vehicles (CAVs)*
According to (*11*) the digitalization of the automotive industry will contribute about US $3.1 trillion in socioeconomic benefits and US $67 billion in business revenue. These benefits aside, CAVs are of interest for the increased efficiency and safety they offer through reducing reliance on humans. CAVs rely heavily on communication in multiple forms, such as vehicle-to-everything communication. Communication is the most cyber-vulnerable area for CAVs. Cyber attacks against CAV communication can restrict the vehicle's internet access and affect vehicle judgment (*20*). Some likely attacks on CAV communication networks include black hole and gray hole attacks, denial of service (DoS) and distributed denial of service attacks, jamming, false message injections, eavesdropping, certificate replication, Sybil attacks, and impersonation (*21*). Such attacks on vehicle autonomy can lead to property damage, injuries, and even loss of life.

**Information Technology and Operational Technology in Transportation Systems**
Cyber risks and cyber-related losses in transportation are rising as new technologies are introduced and dependence on these technologies grows (*5*). In the past, companies relied on the idea that the vulnerabilities of their systems were not widely known and unlikely to be exploited by bad actors. However, this reasoning is no longer valid. Since the digital revolution, many companies have adopted common, low-cost technologies, like Wi-Fi and Ethernet for field devices such as traffic signals, sensors, and dynamic message signs (*22*), which are known to be vulnerable to cyber attacks, such as man-in-the-middle and DoS attacks (*23, 24*).

Information technologies (IT) play a critical role in enabling and managing today's transportation systems by safeguarding and organizing data and information. (*25*). The industry depends on information transmission and reception, making transportation infrastructure reliant on information and communications technology. This reliance increases vulnerability, as bad actors target known weaknesses in these systems (*5*).

Operational technology (OT) can be defined as the computing systems utilized for managing industrial operations and supporting human-to-machine and machine-to-machine interfaces to promote efficiency and automation in systems (*13*). Operational technologies have vulnerabilities resulting from insecure design as well as human factors and configuration issues (*13*). Many OTs, which include supervisory control and data acquisition (SCADA) and distributed control systems (DCSs), are often built with components with known vulnerabilities, e.g., lack of authentication/authorization, insecure defaults, credential mismanagement, and unpatched legacy software (e.g., Windows XP and 2000) (*5, 26*). Some OT systems, such as those in some rail and maritime transportation systems, still rely on legacy systems, which are also vulnerable to cyber attacks (*27, 28*).

IT and OT systems have converged and become more interdependent (*27*). OT systems are often more vulnerable due to reliance on less secure legacy systems. As OT becomes more accessible to cyberspace, it must comply with cybersecurity standards as strict as, or stricter than, those for IT systems (*28*). This is necessary for ports and railways as they adopt new technologies to enhance system intelligence. Implementing these technologies without assessing risks increases vulnerability.

**Supply Chain Cybersecurity**
The transportation industry is comprised of several sub-industries, which conduct business in the same sphere. As a result, supply chain cybersecurity impacts transportation cybersecurity. In transportation, companies rely heavily on suppliers to support their IT and OT systems (*27, 28*). There is a multitude of third-party hardware and software vendors with varying goals and individual functional requirements. It is difficult for companies to ensure that their vendors comply with cybersecurity requirements as very few standards govern vendor cybersecurity. Examples of some of the ways vulnerable vendors can jeopardize the cybersecurity of transportation systems include the introduction of counterfeit hardware (*29*) and hardware trojans (*30*).

**Human Factors**
The expansion of cyberspace in transportation aims to reduce inefficiencies that result from reliance on humans. However, as the industry advances, employees must work with intelligent systems, which requires a certain degree of cybersecurity literacy. Cybersecurity unfamiliarity can benefit attackers (31); for example, 90% of cloud data breaches are reported to be caused by employee errors (*13*). Poorly trained staff pose significant cybersecurity risks, impacting supply chains (*29*). Attackers may also use social engineering with technical attacks to exploit transportation systems (*32*). IBM Security's 2023 X-Force Threat Intelligence Index noted that "phishing was the most common initial access vector in 51% of cases" (*33*). Additionally, transportation systems must be protected from malicious employees, as seen when malware from a crew member's smartphone disrupted a ship's charts, causing a two-day delay (*5*).

**Software Bugs and Updates**
Software bugs and updates also introduce vulnerabilities to transportation cybersecurity. It is estimated that each year, 111 billion lines of new software code are written, harboring billions of vulnerabilities (*13*). Hackers can exploit vulnerabilities in codes to execute various attacks on systems (*13*). For example, software coding errors in the Nissan Leaf were exploited by hackers, allowing them to remotely gain access to the in-car infotainment system and display sensitive driver data (*20*). Systems also become vulnerable when software is not up to date. On the other hand, over-the-air updates also threaten systems due to the risk of malware injections (*21*) and software bugs. Just this year, a software bug in a CrowdStrike update caused major disruptions for numerous airlines and airports, impacting thousands of Windows machines worldwide (*34*).

**POTENTIAL STRATEGIES TO IMPROVE CYBERSECURITY OF THE TRANSPORTATION INDUSTRY**
Previous sections have discussed the transportation industry's vulnerability to cyber attacks. To protect its digital infrastructure from evolving threats, a multifaceted approach is essential. This section discusses potential strategies to improve cybersecurity in transportation considering two aspects: (i) policy and industry and (ii) emerging technologies. These strategies and their benefits are highlighted in **Table 1** and discussed in detail in the following subsections.

**Programmatic and Policy Domain**
Transportation systems face threats across various modes in both the public and private domains. As a result, collaboration throughout the transportation industry is necessary to improve transportation cybersecurity. In this subsection, we present a few collaborative strategies that the transportation industry should address.

## Development of National Standards and Policies

The transportation industry is a vast and diverse web of public and private agencies. Establishing national security standards and policies will heighten industry awareness, enhancing cybersecurity for companies and their collaborators. Some best-practice cybersecurity guidelines are available. For example, NIST SP 800-171 contains standards for controlled unclassified information in nonfederal information systems (*35*). These standards increased in popularity when the US Department of Defense required contractors to comply with security frameworks in DFARS 252.204-7012 (*74*). The NIST Framework for Improving Critical Infrastructure offers cybersecurity standards for critical infrastructure (including transportation). This standard provides insights on risk analysis and risk management for companies (*74*). The ISO/IEC 27000 series, developed by the International Organization for Standardization, includes 60 standards addressing various information security issues, such as ISO/IEC 27018 for cloud computing security and ISO/IEC 27040 for storage security (36, 74). Another important standard is ISO/SAE 21434, the Road Vehicles Cybersecurity Engineering Standard (75). It offers cybersecurity insights for managing risks in the engineering of electrical and electronic systems for road vehicles (*75*). The Open Web Application Security Project (OWASP) provides another security standard called the OWASP Application Security Verification Standard (ASVS). This standard offers an extensive framework for testing web application technical security controls and gives guidelines for security control development and contract specifications (*37*). Implementing national standards and policies such as those mentioned above can equip transportation companies with essential frameworks and guidelines to enhance cybersecurity practices.

## Development of Cybersecurity Testing Strategies

The evolving cyber threats underscore the need for robust testing strategies to protect digital transportation infrastructure by identifying threats, risks, and critical vulnerabilities. An example of this is vulnerability scanning, which identifies weaknesses within systems and evaluates components, such as networks, endpoints, web applications, clouds, and containers, for vulnerabilities using sources like MITRE Corporation's CVE database (*38, 76*). Penetration tests give insight into the potential real-world ramifications of attacks, which exploit vulnerabilities present in systems, as testers simulate attacks on infrastructure, endpoints, applications, clouds, APIs, wireless networks, and users (*38*). Static Application Security Testing (SAST) and secure code reviews are used to detect vulnerabilities within source code, such as injection flaws, buffer overflows, and data exposure, helping companies to identify and mitigate cybersecurity loopholes in their software (*38, 39*). Dynamic Application Security Testing (DAST) is an effective testing strategy that floods production applications with random data to detect vulnerabilities (*38*). Social engineering testing assesses employee vulnerability by employing ethical social engineering services to simulate real-world attack scenarios, such as phishing attempts or the use of physical USB drives containing malware (*38*). Additionally, using threat modeling tools, such as the CAIRIS Threat Modeling Tool, MS Security Development Lifecycle Threat Modeling Tool, and OWASP Threat Dragon, enables organizations to comprehensively analyze security risks by offering a structured representation of relevant information about the application's security and its surrounding environment (*40-42*). Another effective testing strategy involves in-the-loop simulations, such as hardware-in-the-loop (HiL), software-in-the-loop (SiL), and everything-in-the-loop (XiL) simulations, which enable system components to be tested in simulated environments that replicate real-world scenarios (*43*).

**TABLE 1 Potential Strategies to Improve Cybersecurity in the Transportation Industry**

| Programmatic and Policy Domain | | |
|---|---|---|
| **Strategy** | **Benefits** | **Relevant Examples** |
| Development of National Standards and Policies | Heightens awareness of and enforces best practices for cybersecure operations thus potentially strengthening IT and OT security in the era of automation and connectivity | NIST SP 800-171 (*35*), ISO/IEC 27000 series (*36*), OWASP ASVS (*37*) |
| Development of Testing Strategies | Identifies threats, risks, and vulnerabilities in systems before deployment hence potentially preventing attacks including those stemming from human error | Static Application Security Testing (*38,39*), Threat Modeling (*40-42*), In-the-Loop Simulations (*43*) |
| Development of Certification Strategies | Fosters secure business practices, enhancing supply chain security by encouraging partnerships with cybersecure companies | US Department of Defense's use of the Cybersecurity Maturity Model Certification (*44*) |
| Cyber Liability Insurance | Reduces financial losses resulting from cyber attacks | AmTrust Financial, Travelers (*45*) |
| Development of Cybersecurity Workforce | Minimizes cyber incidents resulting from human error | USDOT Intelligent Transportation Systems Professional Capacity Building Program (*46*) |
| Better Reporting of Cyber Incidents | Increases awareness of the current cyber threat landscape, enabling entities to strengthen defenses with updated software patches, testing, certifications, and employee training | Cybersecurity Incident Reporting and Analysis System (*47*), Maritime Cyber Attack Database (*48*) |
| Emerging Technologies Domain* | | |
| **Strategy** | **Benefits** | **Application References** |
| Blockchain with Cryptography | Blockchain ensures data integrity and availability while cryptography ensures confidentiality. | (*49-56*) |
| Zero Trust Architecture (ZTA) and Post-Quantum Cryptography | ZTA protects data through continuous authentication, authorization, and verification, while Post Quantum Cryptography defends against present and quantum threats. | (*57-60*) |
| Confidential Computing and Zero Trust Architecture (ZTA) | Confidential computing protects data in use while ZTA enhances availability. | (*61,62*) |
| Hybrid Hardware-Software Security Approach | A hybrid approach addresses vulnerabilities and threats faced by hardware and software components. | (*63-65*) |
| Satellite-Based Quantum Communication (QC) | This approach helps to address sensor node authentication, trust establishment, and safeguarding sensor data, actuators, and communication channels. | (*66-69*) |

**TABLE 1 Potential Strategies to Improve Cybersecurity in the Transportation Industry (Cont'd)**

| Emerging Technologies Domain* | | |
|---|---|---|
| **Strategy** | **Benefits** | **Application References** |
| Quantum Internet | Quantum internet uses quantum key distribution (QKD), which allows two parties to share a cryptographic key securely. | *(70-73)* |

*Each of these technologies contributes to strengthening IT and OT cybersecurity amid the challenges of increased automation and connectivity.

### Development of Cybersecurity Certification Strategies

As mentioned before, supply chain cybersecurity is linked to overall transportation cybersecurity. Since the transportation industry is interconnected and reliant on suppliers, cybersecurity certification strategies are needed to ensure that companies are adhering to the standards. These standards must be developed with third-party application in mind. Certification schemes are based on standards such as those described in the subsection above and evaluate products and production processes (*77*). An example of this approach is the US Department of Defense's use of the Cybersecurity Maturity Model Certification (*44*).

### Cyber Liability Insurance

Cyber attacks have affected numerous stakeholders in the transportation industry. While the goal is to prevent cybersecurity incidents, some will inevitably occur. The introduction of new technologies brings with it new attack surfaces and vulnerabilities. Companies should have cyber liability insurance to cover and protect their business from financial losses that may result from cybersecurity incidents (*5*). Cyber insurance policies can cover legal services for regulatory compliance, notification expenses for customer data breaches, lost income from network outages, privacy-related lawsuits from employees or customers, and regulatory fines from state or federal agencies (*45*).

### Development of Cybersecurity Workforce

Due to the transportation industry's rapid digitalization, there is a growing need for workers with cybersecurity and resiliency knowledge to protect digital infrastructure. Developing a cybersecurity workforce for transportation should reduce incidents due to human errors. For example, the US Department of Transportation has advanced its cybersecurity workforce through the Intelligent Transportation Systems Professional Capacity Building Program, which educates the workforce on intelligent systems (*46*). The White House's National Cyber Workforce and Education Strategy proposes four pillars to enhance the US cyber workforce and education: "(1) equip every American with foundational cyber skills, (2) transform cyber education, (3) expand and enhance the cyber workforce, and (4) strengthen the federal cyber workforce" (*78*). Therefore, industry and academia should collaborate to create a steady pipeline of cyber-minded individuals.

### Better Reporting of Cyber Incidents

The limited reporting of cybersecurity incidents in transportation poses a significant challenge. Improved reporting can provide industry and academia with a comprehensive view of the evolving threat landscape, as well as insights into common vulnerabilities and attack vectors. To address this, a national database of all known cybersecurity incidents in transportation should be established, enabling those at the forefront of transportation cybersecurity in both industry and academia to access up-to-date information. This access would facilitate the development of more robust cybersecurity protocols for the transportation sector.

Some existing databases help gain insights into the cybersecurity landscape in general. For instance, MITRE's Common Vulnerabilities and Exposures (CVE) database systematically identifies, defines, and catalogs publicly disclosed cybersecurity vulnerabilities, providing consistent descriptions that enable cybersecurity professionals to discuss, prioritize, and address these issues effectively across the industry (*76*). MITRE's Common Weakness Enumeration (CWE) database is an extensive compilation of software and hardware weakness types, created through a community-driven effort to offer precise definitions and detailed information on the effects, behaviors, and exploit mechanisms of each weakness, thereby enhancing comprehension and differentiation for cybersecurity professionals (*79*). The Center for Strategic International Studies has a public database of significant cybersecurity incidents since 2006 (*7*). The European Repository of Cyber Incidents is another public resource, which documents cybersecurity incidents worldwide from the year 2000 to the present day (*80*). University of Maryland's Center for International & Security Studies at Maryland maintains the Cyber Events Database, which gathers data on cyber incidents from 2014 to the present day (*81*).

Some organizations have started addressing the issue of limited reporting in transportation. For example, the European Union Agency for Cybersecurity (ENISA) hosts the Cybersecurity Incident Reporting and Analysis System database, analyzing and visualizing summary reports of cybersecurity incidents affecting European Union (EU) critical service providers since 2012 (*47*). This tool provides statistics on the cyber threat landscape across various critical sectors, including transportation, in the EU. However, individual incidents are not disclosed. Additionally, NHL Stenden University of Applied Sciences maintains the Maritime Cyber Attack Database, which records cybersecurity incidents in maritime transportation using open-source information from 2001 to the present day (*48*). Such databases will only be effective if cyber incident targets commit to a degree of transparency in reporting events.

**Cross-Pollination of Emerging Technology Domains**
In this subsection, we present a technology-focused perspective to address the confidentiality, integrity, and availability requirements of transportation systems and infrastructure. Confidentiality keeps sensitive information unseen by unauthorized individuals, integrity prevents data manipulation during transmission, and availability ensures communication is not disrupted by malicious actors (*82*). While emerging technologies, such as quantum computing, blockchain, and confidential computing, offer significant advantages in meeting these security requirements, we argue that today's information-centric transportation systems would benefit more from their integration and cross-pollination rather than relying on them individually. This section underscores the importance of such hybrid approaches to help prevent cybersecurity failure incidents.

*Blockchain with Cryptography*

Blockchain is a decentralized digital database mechanism that allows users to share information transparently across different nodes or computers within a network. Blockchain stores information in blocks and links them together in a chain, creating a chain of blocks, which cannot be modified without consensus of different nodes from the network (*83*). This immutable chain of blocks serves as a decentralized ledger in which any information shared by an entity is transparent to all other participating entities in the blockchain. Due to its inherently decentralized and consensus-driven nature, blockchain enforces integrity and availability of information. Decentralization is achieved by having each node store a copy of the entire blockchain of which it is a part, ensuring that a single point of failure cannot compromise integrity or availability of the system (*84*). In addition, consensus algorithms, which include proof of work and proof of stake, help validate transactions and newly added blocks, preventing data modification or tampering. However, blockchain implementation cannot enforce confidentiality readily since shared information is transparent to each node within a blockchain; thus arises the necessity to integrate cryptographic schemes with the blockchain technology.

Blockchain leverages cryptographic schemes like hashing, public key cryptography, digital signatures, and Merkle trees, hence ensuring the confidentiality, integrity, and authentication of shared information (*84*). Thus, blockchain with cryptography provides a way to securely store and share data while ensuring confidentiality, integrity, and availability of data.

The transportation industry could benefit from blockchain technology paired with cryptography in numerous ways. Recent studies have shown how blockchain with cryptography can be used to develop multifaceted secure transportation applications, such as parking (*49*), leasing (*50*), and insurance (*51*) management applications, smart fueling (*52*) or charging (*53*) applications, smart frameworks for data sharing and storing among the entities in an Internet of Things environment (*54, 55*), and other payment-associated applications (*56*). However, seamless integration of this technology into transportation applications warrants further exploration to improve the overall security posture of our modern transportation systems and infrastructure.

### Zero Trust Architecture and Post-Quantum Cryptography

Post-Quantum Cryptography (PQC) and Zero Trust Architecture (ZTA) are key security paradigms that together ensure the core principles of cybersecure digital communication: confidentiality, integrity, and availability (*85*). PQC addresses the looming threat of quantum computing by transitioning from current public-key cryptographic algorithms to new ones designed to resist quantum attacks, thus securing data against both present and future vulnerabilities. ZTA, in contrast, is a security architecture that operates under a "never trust, always verify" policy and mandates continuous authentication, authorization, and validation for every user and device, whether inside or outside the organization's network. This ensures that no entity is inherently trusted, and every access request is verified to minimize the risk of unauthorized access and data breaches.

The synergy between ZTA and PQC lies in the principle of explicit verification, which requires rigorous identity management, robust authentication mechanisms, and stringent authorization procedures. These elements demand strong cryptographic algorithms and protocols to ensure cryptographic validation, critical for maintaining confidentiality, integrity, authenticity, availability, and non-repudiation. By leveraging the continuous verification and least privilege principles of ZTA alongside the quantum-resistant cryptographic standards of PQC, our transportation industry can create a security posture that not only mitigates current threats but also anticipates and prepares for future quantum computing challenges. Although few recent studies have explored how vehicular communication networks could benefit from implementing a ZTA or PQC individually (*57–60*), adaptation feasibility, and operational assessment of these technologies' cross-pollination for modern transportation systems remain unexplored.

### Confidential Computing and Zero Trust Architecture

While traditional encryption protects data during storage and transmission over networks, confidential computing secures data in use. Confidential computing employs secure enclaves or trusted execution environments (TEEs), which are isolated regions of a processor that meet the requirements of confidentiality and integrity of the code and data they contain. These TEEs provide a hardware-based, secure area that allows for the execution of sensitive computations without exposing the data to the rest of the system, including the operating system, hypervisor, and hardware. Although confidential computing does not inherently ensure availability, when combined with Zero Trust Architecture (ZTA), it can enhance availability by improving the overall security posture.

ZTA enforces strict access control policies, such as session-based and resource-based access control. However, it still assumes trust relationships among different parts of a processing system or computer. If an attacker gains access to the hardware of such a system, they can perform various attacks, such as side-channel attacks, to reveal sensitive data during processing. Pairing confidential computing with ZTA introduces two critical features: (i) runtime encryption and isolation of workloads, and (ii) remote attestation. The first feature extends ZTA's "never trust" principle to the computing infrastructure by encrypting data in use and ensuring the isolation of different workloads. The second feature requires cryptographic proof that the hardware and software stack of the computing infrastructure can be trusted to process sensitive information.

In an ITS environment, vast amounts of sensitive information are processed and analyzed by numerous entities. For example, CAVs might store and process sensitive information about their users to facilitate various connected and autonomous features, like onboard radio-based electronic payment services. To protect such systems from internal and external attacks, a combination of ZTA and confidential computing is essential. Consequently, in recent years, companies like Intel and Nvidia, manufacturers of high-performance computers, have been focusing on incorporating confidential computing principles into their products (*61*, *62*). Leveraging these as the computing infrastructure while mandating a ZTA would provide a much stronger security posture for future transportation applications.

*Hybrid Hardware-Software Security Approach*

As computing in the IoT environment becomes more decentralized, with distributed and cloud computing services gaining popularity, hardware-related security threats also increase. These threats include attackers gaining physical access to remote compute nodes or the insertion of hardware trojans at any stage of a hardware's lifecycle. Software-based security measures may fail due to these hardware vulnerabilities and threats. The key challenge then becomes maintaining the confidentiality, integrity, and availability of information in such ecosystems. A robust solution to these threats would be adopting a hybrid hardware-software approach to security, rather than focusing on them individually.

A hybrid hardware-software security approach can be implemented in various ways. For example, a hardware-software security co-design approach, proposed by Pesé et al. (*63*), introduced a deployment strategy for automotive embedded firewalls based on optimal hardware-software partitioning. While confidential computing solutions, such as Intel Software Guard Extensions (SGX) and AMD Secure Encrypted Visualization (SEV), offer hardware-based enclaves for secure computing, they still face limitations, such as restricted secure memory and lack of memory integrity protection. To address these issues, researchers have explored different hardware-software co-design mechanisms, such as using lightweight software to store the logic of enclave mechanisms (*64*) and implementing hybrid hardware-software remote attestation approaches (*65*). These hybrid methods can provide more robust protection of confidentiality, integrity, and availability for transportation-related information than either software-only or hardware-only solutions. However, in a dynamic environment like transportation, implementing these approaches requires comprehensive testing to ensure they meet critical application requirements—a topic that still requires further research.

*Satellite-based Quantum Communication (QC)*

Satellite-based Quantum Communication (QC) can enhance the cybersecurity and resiliency of connected and automated transportation systems. Researchers are leveraging satellite-based QC to address several key challenges related to sensor networks, including sensor node authentication, trust establishment, and safeguarding sensor data, actuators, and communication channels. The use of quantum key distribution (QKD) in satellite communication aims to provide a higher level of security for data transfer by utilizing the principles of quantum mechanics. This approach can ensure secure key exchange over long distances (*86*). These systems implement high-speed quantum random number generation and polarization encoding, coupled with precise satellite tracking and pointing mechanisms, to ensure reliable and efficient key exchange despite dynamic satellite movements and operational constraints. Recent developments highlight the potential of Low Earth Orbit (LEO) satellites to achieve real-time QKD using satellite communication (*66*). This progress has increased by high-speed data transfer and precise optical tracking mechanisms. These advancements inspire the practical implementation of secure QC networks on a global scale, although the range of LEO is limited, and could enhance the resilience of navigation systems by effectively mitigating operational challenges, such as signal fidelity and atmospheric interference.

Satellites deployed in Medium Earth Orbit (MEO), like in the Global Navigation Satellite System (GNSS) constellation, significantly extend secure communication ranges. This wide range of quantum channels is required for establishing QKD protocols over longer distances, secure GNSS against threats like spoofing and man-in-the-middle (MiM) attacks. For instance, experiments utilizing GLONASS satellites equipped with retroreflector arrays have validated the transmission of single photons over extensive distances, showcasing advancements in QC technologies (*67*). These initiatives pave the way for satellite-based quantum networks, enhancing security protocols across sensor nodes, high-mobility vehicles, and transportation communication interfaces. The integration of high-orbit satellites, such as those in GNSS, offers a pathway for establishing global-scale quantum networks. Experimental validations utilizing satellite-based QC have been achieved in secure key distribution using entanglement-based protocols over distances exceeding 1,120 kilometers (*68*). These efforts highlight the potential for deploying QC-enabled satellite networks to enhance cybersecurity resilience across transportation cyber-physical systems by leveraging advanced quantum cryptographic protocols like BB84 QKD, optimized for satellite communication channels (*69*). In addition, the evolution towards post-quantum secure authentication protocols implemented in satellite communication systems initiates measures against emerging quantum computing threats.

*Quantum Internet*

The quantum internet addresses cybersecurity and resiliency challenges by leveraging quantum mechanics principles like superposition and entanglement. An important application of the quantum internet is QKD, which allows two parties to share a cryptographic key securely. Any attempt to eavesdrop on the key would disturb the quantum states, alerting the communicating parties to the presence of an intruder, thereby making the communication theoretically unhackable (*70*). Additionally, the no-cloning theorem prevents the duplication of quantum information, further securing communications. The integration of quantum and classical networks facilitates secure and reliable transmissions through protocols, such as QKD, which generates encryption keys based on quantum principles. Research in quantum teleportation and entanglement distribution is critical for developing the Quantum Internet, ensuring secure and efficient data transmission by mitigating issues related to quantum decoherence and transmission fidelity (*71*). A study shows the advancement of trusted computing platform standards, ensuring robust and resilient network infrastructure capable of withstanding quantum-specific threats (*72*). Another study provides a comprehensive review of the potential applications and underlying concepts of a long-range QC network, or quantum internet (*73*). It discusses the secure transmission of classical and quantum information, highlighting the principles of error correction, teleportation, and the use of quantum repeaters.

**CONCLUSIONS**

As cyberspace in the transportation industry continues to expand, attacks on digital infrastructure are becoming increasingly serious and problematic. This study highlighted several recent cyberattack incidents affecting transportation systems, illustrating the diverse and evolving threat landscape. It also underscored how factors, such as increased connectivity and automation, supply chain vulnerabilities, inadequately trained personnel, software flaws, and challenges associated with software updates, contribute to the rise in cyber vulnerabilities. With the threat landscape in transportation growing in complexity, there is an urgent need for robust and collaborative cybersecurity measures to secure digital infrastructure.

Key strategies to improve the cybersecurity climate in the transportation industry include establishing national cybersecurity standards and policies, implementing rigorous cybersecurity testing protocols, adopting certification strategies, developing a cybersecurity workforce, improving cyber incident reporting mechanisms, and mandating cyber liability insurance. These strategies can help secure the industry's digital infrastructure in an ever-evolving threat landscape, enhancing the safety and security of transportation systems worldwide. Additionally, the integration of emerging technologies to produce hybrid cybersecurity solutions offers substantial advantages in addressing the confidentiality, integrity, and availability requirements of transportation systems. This paper explores the potential cross-pollination of technologies, such as blockchain with cryptography, zero trust architecture and post-quantum cryptography, confidential computing with zero trust architecture, hybrid hardware-software security, satellite-based quantum computing, and quantum internet. These hybrid technologies can significantly enhance the cybersecurity of transportation systems. Future research and development in these areas are needed for advancing cybersecurity and resiliency of transportation systems.

## ACKNOWLEDGMENTS

This work is based upon the work supported by the National Center for Transportation Cybersecurity and Resiliency (TraCR) (a U.S. Department of Transportation National University Transportation Center) headquartered at Clemson University, Clemson, South Carolina, USA. Any opinions, findings, conclusions, and recommendations expressed in this material are those of the author(s) and do not necessarily reflect the views of TraCR, and the U.S. Government assumes no liability for the contents or use thereof. Chat GPT 3.5 and 4-o were used only to edit the text in this manuscript.

## AUTHOR CONTRIBUTIONS

The authors confirm contribution to the paper as follows: study conception and design: M. Chowdhury, O. Thomas; data collection: O. Thomas; analysis and interpretation of results: O. Thomas, J. Tine, S. Salek; draft manuscript preparation: O. Thomas, J. Tine, S. Salek, M. Chowdhury, M. Rahman, T. Hockstad. All authors reviewed the results and approved the final version of the manuscript.

**REFERENCES**

1. Cost of a Data Breach 2024 | IBM. https://www.ibm.com/reports/data-breach. Accessed Jul. 31, 2024.

2. Hope, A. Ransomware Attack Linked to Permanent Shut Down of Illinois Hospital St. Margaret's Health in Spring Valley. CPO Magazine, Jun 20, 2023.

3. The Attack on Colonial Pipeline: What We've Learned & What We've Done Over the Past Two Years | CISA. https://www.cisa.gov/news-events/news/attack-colonial-pipeline-what-weve-learned-what-weve-done-over-past-two-years. Accessed Jul. 31, 2024.

4. Bowcut, S. Cybersecurity in the transportation industry, 2023. Available at https://cybersecurityguide.org/industries/transportation/#:~:text=According%20to%20Cybertalk.org%2C%20between,the%20brunt%20of%20this%20trend. Last Accessed August 31st, 2023.

5. Tonn, G., J. P. Kesan, L. Zhang, and J. Czajkowski. Cyber Risk and Insurance for Transportation Infrastructure. *Transport Policy*, Vol. 79, 2019, pp. 103–114. https://doi.org/10.1016/j.tranpol.2019.04.019.

6. Ćosić, J. & Mihailescu, M.I. ENISA Threat Landscape: Transport Sector, 2023.

7. Significant Cyber Incidents | Strategic Technologies Program | Center for Strategic and International Studies (CSIS). https://www.csis.org/programs/strategic-technologies-program/significant-cyber-incidents. Accessed May 29, 2024.

8. Ukwandu, E., M. A. Ben-Farah, H. Hindy, M. Bures, R. Atkinson, C. Tachtatzis, I. Andonovic, and X. Bellekens. Cyber-Security Challenges in Aviation Industry: A Review of Current and Future Trends. *Information*, Vol. 13, No. 3, 2022, p. 146. https://doi.org/10.3390/info13030146.

9. Meland, P. H., K. Bernsmed, E. Wille, Ø. J. Rødseth, and D. A. Nesheim. A Retrospective Analysis of Maritime Cyber Security Incidents. *519-530*, 2021. https://doi.org/10.12716/1001.15.03.04.

10. Callahan, J. Transportation Cybersecurity Incidents. https://ati.ua.edu/wp-content/uploads/2021/05/72.pdf. Accessed May 29, 2024.

11. IBM Security. Cost of a Data Breach Report 2023, 2023.

12. International Association of Ports and Harbors (IAPH). Cybersecurity Guidelines for Ports and Port Facilities, 2021. https://sustainableworldports.org/wp-content/uploads/IAPH-Cybersecurity-Guidelines-version-1_0.pdf. Accessed May 29, 2024.

13. Lehto, M., and P. Neittaanmäki. *Cyber Security*. Springer Nature, 2022.

14. Greig, J. Major Airline Technology Provider Accelya Attacked by Ransomware Group. https://therecord.media/major-airline-technology-provider-accelya-attacked-by-ransomware-group. Accessed Jul. 23, 2024.

15. Sutton, H. I. Positions of Two NATO Ships Were Falsified Near Russian Black Sea Naval Base. USNI News, Jun 21, 2021.

16. Tabak, N. Ransomware Attack Hits Rail Freight Operator OmniTRAX. *FreightWaves*. https://www.freightwaves.com/news/ransomware-attack-hits-short-line-rail-operator-omnitrax. Accessed Jul. 20, 2024.

17. Osorio, N. Russians in Novosibirsk Forced To Pound Pavements As Team OneFist Paralyzes Traffic. *International Business Times*. https://www.ibtimes.com/russians-novosibirsk-forced-pound-pavements-team-onefist-paralyzes-traffic-exclusive-3611628. Accessed May 27, 2024.

18. Oregon Department of Transportation : MOVEit Data Breach : Oregon Driver & Motor Vehicle Services : State of Oregon. https://www.oregon.gov/odot/dmv/pages/data_breach.aspx. Accessed Aug. 1, 2024.

19. Demertzi, V., S. Demertzis, and K. Demertzis. An Overview of Cyber Threats, Attacks and Countermeasures on the Primary Domains of Smart Cities. *Applied Sciences*, Vol. 13, No. 2, 2023, p. 790. https://doi.org/10.3390/app13020790.

20. Khan, S. K., N. Shiwakoti, P. Stasinopoulos, and Y. Chen. Cyber attacks in the next-Generation Cars, Mitigation Techniques, Anticipated Readiness and Future Directions. *Accident Analysis & Prevention*, Vol. 148, 2020, p. 105837. https://doi.org/10.1016/j.aap.2020.105837.

21. Chowdhury, M., M. Islam, and Z. Khan. Security of Connected and Automated Vehicles. *The Bridge, National Academy of Engineering,* Vol. 39, No. 3, 2019, pp. 46-56.

22. Hou, Y., K. Collins, M. V. Wart, Y. Hou, K. Collins, and M. V. Wart. Intersection Management, Cybersecurity, and Local Government: ITS Applications, Critical Issues, and Regulatory Schemes. In *Smart Mobility - Recent Advances, New Perspectives and Applications*, IntechOpen.

23. Ingers, J. and Sjöblom, J. IoT Penetration Testing: Examining the Cybersecurity of Connected Vehicles, 2022. https://www.diva-portal.org/smash/get/diva2:1703901/FULLTEXT01.pdf Accessed May 29, 2024.

24. El-Rewini, Z., K. Sadatsharan, D. F. Selvaraj, S. J. Plathottam, and P. Ranganathan. Cybersecurity Challenges in Vehicular Communications. *Vehicular Communications*, Vol. 23, 2020, p. 100214. https://doi.org/10.1016/j.vehcom.2019.100214.

25. Sechi, F. (2023). Critical Convergence for Enhanced Safety: A Literature Review on Integrated Cybersecurity Strategies for Information Technology and Operational Technology Systems Within Critical Infrastructure. *Proceedings of the 33rd European Safety and Reliability Conference (ESREL 2023)*.

26. Silverman, D., Y.-H. Hu, and M. Hoppa. A Study on Vulnerabilities and Threats to SCADA Devices. *Journal of The Colloquium for Information Systems Security Education*, Vol. 7, No. 1, 2020, pp. 8–15.

27. Drougkas, A., Sarri, A., Kyranoudi, P., and Zisi, A. Port Cybersecurity - Good Practices for Cybersecurity in the Maritime Sector. *ENISA*. https://www.enisa.europa.eu/publications/port-cybersecurity-good-practices-for-cybersecurity-in-the-maritime-sector. Accessed May 29, 2024.

28. Liveri, D., Theocharidou, M., and Naydenov, R. Railway Cybersecurity. *ENISA*. https://www.enisa.europa.eu/publications/railway-cybersecurity. Accessed May 29, 2024.

29. Afenyo, M., and L. D. Caesar. Maritime Cybersecurity Threats: Gaps and Directions for Future Research. *Ocean & Coastal Management*, Vol. 236, 2023, p. 106493. https://doi.org/10.1016/j.ocecoaman.2023.106493.

30. Tine, J. M., S.-N. Puspa, R. Majumdar, G. Comert, M. Chowdhury, and Y. Lao. Threats of Trojan Incursion in Transportation Hardware. Presented at the 2023 IEEE International Automated Vehicle Validation Conference (IAVVC), 2023.

31. Akpan, F., G. Bendiab, S. Shiaeles, S. Karamperidis, and M. Michaloliakos. Cybersecurity Challenges in the Maritime Sector. *Network*, Vol. 2, No. 1, 2022, pp. 123–138. https://doi.org/10.3390/network2010009.

32. Unwin, D., and L. Sanzogni. Railway Cyber Safety: An Intelligent Threat Perspective. *Proceedings of the Institution of Mechanical Engineers, Part F: Journal of Rail and Rapid Transit*, Vol. 236, No. 1, 2022, pp. 26–34. https://doi.org/10.1177/09544097211000518.

33. IBM Security. X-Force Threat Intelligence Index 2023, 2023.

34. Warren, T. Major Windows BSOD Issue Hits Banks, Airlines, and TV Broadcasters. *The Verge*. https://www.theverge.com/2024/7/19/24201717/windows-bsod-crowdstrike-outage-issue. Accessed Jul. 29, 2024.

35. Ross, R. *Protecting Controlled Unclassified Information in Nonfederal Systems and Organizations*. Publication NIST SP 800-171r3. National Institute of Standards and Technology, Gaithersburg, MD, 2024, p. NIST SP 800-171r3.

36. ISO - ISO/IEC 27000 Family — Information Security Management. *ISO*. https://www.iso.org/standard/iso-iec-27000-family. Accessed May 27, 2024.

37. OWASP Application Security Verification Standard | OWASP Foundation. https://owasp.org/www-project-application-security-verification-standard/. Accessed May 27, 2024.

38. Patel, H. Staying a Step Ahead: A Guide to Cybersecurity Testing Methods. WPG Consulting, Sep 06, 2023.

39. Top 5 Cybersecurity Testing Methods. *Kirbtech.* Nov 07, 2022.

40. CAIRIS. https://cairis.org/. Accessed May 27, 2024.

41. Microsoft Threat Modeling Tool Overview - Azure. https://learn.microsoft.com/en-us/azure/security/develop/threat-modeling-tool. Accessed May 27, 2024.

42. OWASP Threat Dragon | OWASP Foundation. https://owasp.org/www-project-threat-dragon/. Accessed May 27, 2024.

43. How XiL Testing Can Revolutionise the Automotive Industry | AB Dynamics. https://www.abdynamics.com/en/blog/2023/xil-testing-the-end-of-prototype-vehicle-testing. Accessed May 27, 2024.

44. CMMC Model. https://dodcio.defense.gov/CMMC/Model/. Accessed May 27, 2024.

45. Tretina, K. The 5 Best Cyber Insurance Companies of 2024. *Investopedia*. https://www.investopedia.com/best-cyber-insurance-5069694. Accessed May 27, 2024.

46. USDOT ITS Research - ITS Cybersecurity Workforce Development. https://www.its.dot.gov/research_areas/cybersecurity/workforce.htm. Accessed May 27, 2024.

47. ENISA. Incident Reporting. *CIRAS*. https://ciras.enisa.europa.eu/ciras-consolidated-reporting. Accessed Jun. 11, 2024.

48. Maritime Cyber Attack Database (MCAD) | NHL Stenden University of Applied Sciences. https://www.nhlstenden.com/en/maritime-cyber attack-database. Accessed Jun. 11, 2024.

49. Jabbar, R., M. Krichen, M. Shinoy, M. Kharbeche, N. Fetais, and K. Barkaoui. A Model-Based and Resource-Aware Testing Framework for Parking System Payment Using Blockchain. Presented at the 2020 International Wireless Communications and Mobile Computing (IWCMC), 2020.

50. Obour Agyekum, K. O.-B., Q. Xia, E. Boateng Sifah, S. Amofa, K. Nketia Acheampong, J. Gao, R. Chen, H. Xia, J. C. Gee, X. Du, and M. Guizani. V-Chain: A Blockchain-Based Car Lease Platform. Presented at the 2018 IEEE International Conference on Internet of Things (iThings) and IEEE Green Computing and Communications (GreenCom) and IEEE Cyber, Physical and Social Computing (CPSCom) and IEEE Smart Data (SmartData), 2018.

51. Li, Z., Z. Xiao, Q. Xu, E. Sotthiwat, R. S. Mong Goh, and X. Liang. Blockchain and IoT Data Analytics for Fine-Grained Transportation Insurance. Presented at the 2018 IEEE 24th International Conference on Parallel and Distributed Systems (ICPADS), 2018.

52. Jamil, F., O. Cheikhrouhou, H. Jamil, A. Koubaa, A. Derhab, and M. A. Ferrag. PetroBlock: A Blockchain-Based Payment Mechanism for Fueling Smart Vehicles. *Applied Sciences*, Vol. 11, No. 7, 2021, p. 3055. https://doi.org/10.3390/app11073055.

53. Su, Z., Y. Wang, Q. Xu, M. Fei, Y.-C. Tian, and N. Zhang. A Secure Charging Scheme for Electric Vehicles With Smart Communities in Energy Blockchain. *IEEE Internet of Things Journal*, Vol. 6, No. 3, 2019, pp. 4601–4613. https://doi.org/10.1109/JIOT.2018.2869297.

54. Jiang, T., H. Fang, and H. Wang. Blockchain-Based Internet of Vehicles: Distributed Network Architecture and Performance Analysis. *IEEE Internet of Things Journal*, Vol. 6, No. 3, 2019, pp. 4640–4649. https://doi.org/10.1109/JIOT.2018.2874398.

55. Yang, Z., K. Yang, L. Lei, K. Zheng, and V. C. M. Leung. Blockchain-Based Decentralized Trust Management in Vehicular Networks. *IEEE Internet of Things Journal*, Vol. 6, No. 2, 2019, pp. 1495–1505. https://doi.org/10.1109/JIOT.2018.2836144.

56. Jabbar, R., N. Fetais, M. Kharbeche, M. Krichen, K. Barkaoui, and M. Shinoy. Blockchain for the Internet of Vehicles: How to Use Blockchain to Secure Vehicle-to-Everything (V2X) Communication and Payment? *IEEE Sensors Journal*, Vol. 21, No. 14, 2021, pp. 15807–15823. https://doi.org/10.1109/JSEN.2021.3062219.

57. Anderson, J., Q. Huang, L. Cheng, and H. Hu. A Zero-Trust Architecture for Connected and Autonomous Vehicles. *IEEE Internet Computing*, Vol. 27, No. 5, 2023, pp. 7–14. https://doi.org/10.1109/MIC.2023.3304893.

58. Shipman, M. E., N. Millwater, K. Owens, and S. Smith. A Zero Trust Architecture for Automotive Networks. 2024. https://doi.org/10.4271/2024-01-2793.

59. Malina, L., P. Dzurenda, S. Ricci, J. Hajny, G. Srivastava, R. Matulevičius, A.-A. O. Affia, M. Laurent, N. H. Sultan, and Q. Tang. Post-Quantum Era Privacy Protection for Intelligent Infrastructures. *IEEE Access*, Vol. 9, 2021, pp. 36038–36077. https://doi.org/10.1109/ACCESS.2021.3062201.

60. Khalid, H., S. J. Hashim, F. Hashim, W. A. M. Al-Jawher, M. A. Chaudhary, and H. H. M. Altarturi. RAVEN: Robust Anonymous Vehicular End-to-End Encryption and Efficient Mutual Authentication for Post-Quantum Intelligent Transportation Systems. *IEEE Transactions on Intelligent Transportation Systems*, 2024, pp. 1–13. https://doi.org/10.1109/TITS.2024.3416060.

61. Confidential Computing Solutions. *Intel*. https://www.intel.com/content/www/us/en/security/confidential-computing.html. Accessed Jul. 26, 2024.

62. Apsey, E., P. Rogers, M. O'Connor, and R. Nertney. Confidential Computing on NVIDIA H100 GPUs for Secure and Trustworthy AI. *NVIDIA Technical Blog*. https://developer.nvidia.com/blog/confidential-computing-on-h100-gpus-for-secure-and-trustworthy-ai/. Accessed Jul. 26, 2024.

63. Pesé, M. D., K. Schmidt, and H. Zweck. *Hardware/Software Co-Design of an Automotive Embedded Firewall*. Publication 2017-01–1659. SAE International, Warrendale, PA, 2017.

64. Enclavisor: A Hardware-Software Co-Design for Enclaves on Untrusted Cloud | IEEE Journals & Magazine | IEEE Xplore. https://ieeexplore.ieee.org/abstract/document/9178442?casa_token=fonaQ-M4fVEAAAAA:BuxjB5vwFWiaRzsqcuoJNoPsQJ5w0AERiT-ONCTyYZzvoQFHKBRPabYFLNci2RUDB4kTMuHIFg. Accessed Jul. 26, 2024.

65. Nunes, I. D. O., K. Eldefrawy, N. Rattanavipanon, M. Steiner, and G. Tsudik. VRASED: A Verified Hardware/Software Co-Design for Remote Attestation. Presented at the 28th USENIX Security Symposium (USENIX Security 19), 2019.

66. Roger, T., Singh, R., Perumangatt, C., Marangon, D.G., Sanzaro, M., Smith, P.R., Woodward, R.I. and Shields, A.J., 2023. Real-time gigahertz free-space quantum key distribution within an emulated satellite overpass. *Science Advances*, *9*(48), p.eadj5873.

67. Calderaro, L., Agnesi, C., Dequal, D., Vedovato, F., Schiavon, M., Santamato, A., Luceri, V., Bianco, G., Vallone, G. and Villoresi, P., 2018. Towards quantum communication from global navigation satellite system. *Quantum Science and Technology*, *4*(1), p.015012.

68. Yin, J., Li, Y.H., Liao, S.K., Yang, M., Cao, Y., Zhang, L., Ren, J.G., Cai, W.Q., Liu, W.Y., Li, S.L. and Shu, R., 2020. Entanglement-based secure quantum cryptography over 1,120 kilometres. *Nature*, *582*(7813), pp.501-505.

69. Hosseinidehaj, N. and Malaney, R., 2016. CV-MDI quantum key distribution via satellite. *arXiv preprint arXiv:1605.05445*.

70. The Quantum Internet, Explained | Pritzker School of Molecular Engineering | The University of Chicago. https://pme.uchicago.edu/news/quantum-internet-explained-0. Accessed Jul. 31, 2024.

71. Valivarthi, R., Davis, S.I., Peña, C., Xie, S., Lauk, N., Narváez, L., Allmaras, J.P., Beyer, A.D., Gim, Y., Hussein, M. and Iskander, G., 2020. Teleportation systems toward a quantum internet. *PRX Quantum*, *1*(2), p.020317.

72. Cacciapuoti, A.S., Caleffi, M., Tafuri, F., Cataliotti, F.S., Gherardini, S. and Bianchi, G., 2019. Quantum internet: Networking challenges in distributed quantum computing. *IEEE Network*, *34*(1), pp.137-143.

73. Dür, W., Lamprecht, R. and Heusler, S., 2017. Towards a quantum internet. *European Journal of Physics*, *38*(4), p.043001.

74. Kirvan, P. Top 12 IT Security Frameworks and Standards Explained | TechTarget. *Security*. https://www.techtarget.com/searchsecurity/tip/IT-security-frameworks-and-standards-Choosing-the-right-one. Accessed May 27, 2024

75. ISO/SAE 21434:2021(En), Road Vehicles — Cybersecurity Engineering. https://www.iso.org/obp/ui/en/#iso:std:iso-sae:21434:ed-1:v1:en. Accessed May 27, 2024.

76. CVE Website. https://www.cve.org/. Accessed May 27, 2024.

77. What's the Difference between Testing and Certification? *Solar Rating & Certification Corporation*. https://solar-rating.org/faq/whats-the-difference-between-testing-and-certification/. Accessed May 27, 2024.

78. National Cyber Workforce and Education Strategy. *Office of the National Cyber Director, Executive Office of the President.* https://www.whitehouse.gov/wp-content/uploads/2023/07/NCWES-2023.07.31.pdf. Accessed May 27, 2024.

79. CWE - CWE List Version 4.14. https://cwe.mitre.org/data/index.html. Accessed May 27, 2024.

80. The European Repository of Cyber Incidents. *EuRepoC: European Repository of Cyber Incidents*. https://eurepoc.eu/. Accessed Jun. 11, 2024.

81. Cyber Events Database | Center for International and Security Studies at Maryland. https://cissm.umd.edu/cyber-events-database. Accessed Jun. 11, 2024.

82. A Framework to Understand Cybersecurity. *NAE Website*. https://nae.edu/216553/A-Framework-to-Understand-Cybersecurity. Accessed Aug. 1, 2024.

83. Zheng, Z., S. Xie, H. N. Dai, X. Chen, and H. Wang. Blockchain Challenges and Opportunities: A Survey. International Journal of Web and Grid Services, Vol. 14, No. 4, 2018, p. 352. https://doi.org/10.1504/IJWGS.2018.095647.

84. Bachchas, K. S. Deep Dive into Blockchain Security: Vulnerabilities and Protective Measures. https://levelblue.com/blogs/security-essentials/deep-dive-into-blockchain-security-vulnerabilities-and-protective-measures. Accessed Nov. 27, 2024

85. Post-Quantum Cryptography & Zero Trust | SandboxAQ. https://www.sandboxaq.com/post/bridging-post-quantum-cryptography-and-zero-trust-architecture. Accessed Jul. 26, 2024.

86. Chou, H.F., Ha, V.N., Al-Hraishawi, H., Garces-Socarras, L.M., Gonzalez-Rios, J.L., Merlano-Duncan, J.C. and Chatzinotas, S., *Satellite-based Quantum Network: Security and Challenges over Atmospheric Channel* (No. arXiv: 2308.00011).